**From MVPs to pivots: a hypothesis-driven journey of two software startups**

Dron Khanna[1], Anh Nguyen-Duc[2], and Xiaofeng Wang[1]

[1] Free University of Bozen-Bolzano, Bolzano 39100, Italy, dron.khanna@unibz.it, xiaofeng.wang@unibz.it

[2] University of Southeast Norway, 3800 Bø i Telemark, Norway anh.nguyen.duc@usn.no

**Abstract**. Software startups have emerged as an interesting multiper- spective research area. Inspired by Lean Startup, a startup journey can be viewed as a series of experiments that validate a set of business hy- potheses an entrepreneurial team make explicitly or inexplicitly about their startup. It is little known about how startups evolve through busi- ness hypothesis testing. This study proposes a novel approach to look at the startup evolution as a Minimum Viable Product(MVP) creat- ing process. We identified relationships among business hypotheses and MVPs via ethnography and post-mortem analysis in two software star- tups. We observe that the relationship between hypotheses and MVPs is incomplete and non-linear in these two startups. We also find that entrepreneurs do learn from testing their hypotheses. However, there are hypotheses not tested by MVPs and vice versa, MVPs not related to any business hypothesis. The approach we proposed visualizes the flow of entrepreneurial knowledge across pivots via MVPs.

**Keywords**: Software startup, Lean startup, Entrepreneurial journey, Minimum Viable Product, Pivot

## 1 Introduction

The software industry has witnessed a growing trend of the development of soft- ware products by small teams of people with limited resource and little operating history. Despite this global movement of high-tech entrepreneurship, the major- ity of software startups fail within two years of their creation, primarily due to self-destruction rather than competition [1]. The number will be much higher when counting startup teams which have not reached the launching milestone. It is known that there is no common recipe for entrepreneurs to be successful. It is difficult to frame successes and failure from startups [2], as each startup will have a unique evolution path depending on an abundant amount of context factors. Lean startup, a common methodology among entrepreneur, emphasizes the role of validating business ideas via building MVPs. It is also common that a pivot occurs after a series of MVPs are created [3,4]. Such a startup journey is also an artefact-creating process, given that major milestones for startups (namely: pitching events, first paid customer and fund-raising) tight to certain artefacts. Entrepreneurship research provides a grounded foundation that startup is an emergent sequence of events, in which an event is both, path dependent on prior processes and contingent on contemporaneous processes [1, 5–7].

While it is useful for an entrepreneur to view entrepreneurial development from an MVP-creating process perspective, it is more important for them to know what they can learn from their MVPs. Ries mentions the Build-Measure-Learn circle in his method [8]. The concept of the loop explains that build stage is based on the hypothesis formulated by an entrepreneur. In order to test the hypothesis, an experiment has to be configured. Learning is intended during the testing of hypothesis [9]. Therefore, this loop could also be regarded interpreted as a traditional scientific hypothesis-metric-experiment loop. The cycle that starts with the hypothesis and ends with a prototype to test the hypothesis. While exercising the loop, the earlier a startup realizes a hypothesis is wrong, the quicker it should be updated and retested [9]. However, the cycle does not directly imply what software entrepreneur actually learn from their previous experience embedded in MVPs. Software startup teams are excessively focused on the developing a better software solution and delivering a prototype to its customer. Individuals exercising so many experiments to win the software development timeline, often neglect the learning involved in software startups [10]. The objective of this study is to understand the entrepreneurial learning from an MVP-creation process. We assume that entrepreneur has predetermined business ideas, which are formed as a hypothesis, that is validated by building MVPs. Therefore, adopting MVP as the unit of analysis, our research questions are RQ1: Do entrepreneur learn from formulated hypotheses for their business and product? RQ2: Are their corresponding MVPs for a formulated hypothesis? The study is organized as follows: Section 2 presents a background about startup development and entrepreneurial artefacts. Section 3 describes our study design, case description, data collection and data analysis. Section 4 presents the entrepreneurial journey of two software startups: Startuppuccino and MUML AS. Finally, Section 5 presents the discussion and concludes the paper.

## 2 Background and Related Work

To explore our research questions, we articulate two theoretical fields: startup development and entrepreneurial artefacts as illustrated in Figure 1. On the grounds of software engineering, a startup doing experiments contributes with knowledge on software development process, techniques and their outcomes. The procedure to carry out experimentation helps the startup team to better predict, understand and develop the software development process [11].

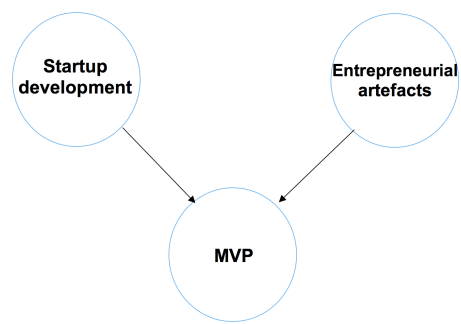

Fig. 1. Theoretical aspects of MVP's

### 2.1 Startup Development

Lean Startup [8] as a methodology for entrepreneurship has become increasingly popular in the past several years, evidenced by dedicated conferences and global Lean Startup meet-ups. As a result, it starts to enter entrepreneurship education programs as the main topic too. The Lean Startup approach was inspired by the lean concepts of focusing on the efforts that create value for customers and eliminating waste during entrepreneurial processes [8]. However, since the customers are often unknown, what customers could perceive as value is also unknown. Therefore, entrepreneurs should get out of the building to involve the customers since day one [12]. Lean Startup advocates to build the product iteratively and deliver to the market as quickly as possible for earlier feedback [8]. Lean Startup is essentially a hypothesis-driven approach [13] which bases entrepreneurial decisions on evidence and validated learning. To capture customer value, an entrepreneur should start a feedback loop that turns an idea into a product, learning whether to pivot or persevere. This can be done by developing an MVP using agile methods to collect customer feedback about the product [8]. The feedback becomes the input to improve the product and validate the hypothesis. As a result, the startup might pursue new directions of the business or continue and scale it [14]. Figure 2 is a high-level representation of the Lean Startup methodology. Pivots in software startups are common to occur and discussed by various scholars.

Fig. 2. Lean Startup Process Model [14]

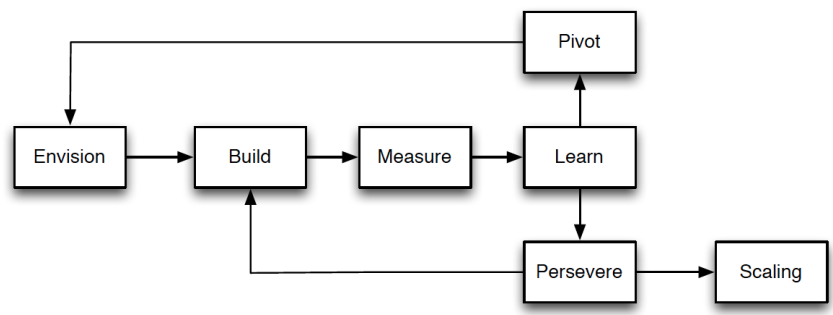

According to Ries [8], it is a kind of change done to validate the startup hypothesis about a product, business model and the engine of growth. Bajwa et al. in their study refer to various different types of pivots that can happen in startups: Zoom-in, Zoom-out, Customer Segment, Customer need, Platform, Business Architecture, Value Capture, Engine of Growth, Channel, Technology, Complete and Side project [4]. A startup journey can be seen as a process of creating entrepreneurial artefacts [15]. According to the science of artificial, one of the schools of theory adopted in entrepreneurship research [16], an artefact is defined as an interface between the internal team and its surrounding environment. MVP is one type of artefact created as a result of the entrepreneurial process. As a core concept of Lean startup [8], MVP is a version of a new product which allows a team to collect the maximum amount of know-how about customers with the least effort [8]. Eric Ries listed several types of MVPs, for example, an explainer video, a landing page, a wire-frame, and a single feature prototype [8]. In Software Engineering context, Nguyen Duc et al. discussed the throw-away prototype and the evolutionary prototype as an MVP [17]. MVP is also considered as a type of boundary object in startup context [3].

**2.2 Theoretical Model of Startup Evolution**

Based on the Build-Measure-Learn approach, hypothesis about both product and customer should be formed and validated using MVPs [8]. The loop repeats and moves forward, from problem-solution space to product-market space and eventually to scaling. Lindgren and Mu ̈nch present a study about experiment- driven product development in the startup context. The authors describe the product development as a series of linear increment of experiments [18]. Fager- holm et al. propose a framework for the continuous experiment which includes the elements of the lean startup [19]. This type of experiment points out the importance of continuous testing in order to support the development process to achieve the high-end product. Continuous in this context refers to running many iterations of Build-Measure-Learn feedback loop. In addition to whisking the experiment Fagerholm et al. provides the description of required artefacts, tasks and roles [19, 18]. This experiment-driven process facilitates the development of MVP or minimum viable features (MVF) and supports the plan, implementation and analysis of experiments. Holmstr ̈om et al. study describes the Hypothesis Experiment Data-Driven Development (HYPEX) model which helps to blend the experiments with the customer in the software development process. The HYPEX model aims at reducing the customer feedback loop. Hence this leads to less development pressure in the software development process. Similar to the approaches mentioned earlier Nguyen et al. represents the evolution of star- tups via double loop model of sense-making [20]. We formed a process-based framework to realize the entrepreneurial process as in Figure 3.

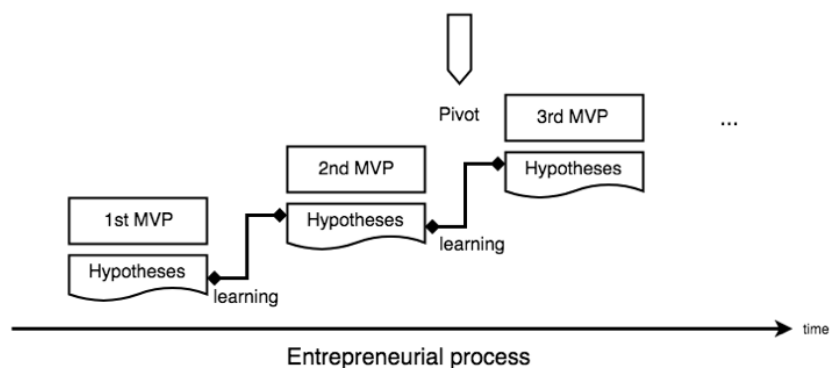

Fig. 3. Hypothetical process of artefact-driven startup evolution

## 3 Research Approach

This section describes the research methodology adopted to study our cases. Given startups are a dynamic and multi-influenced environment, our initial plan was to conduct an exploratory case study. Further, in the research process, our data was dominated by participant observations due to the fact that all of the paper authors were heavily involved in the startup cases. This motivated us to conduct a tailor ethnography study [21]. Ethnography derives from traditional anthropology aiming at telling a credible, rigorous, and authentic story, giving voice to people in their local context [22]. The central focus of ethnography is to provide rich, holistic insights into people's views and actions, as well as the scenario where they behave, through the collection of detailed observations and interviews [23]. There have been some attempts to adopt ethnography in software engineering context [24]. In this type of study, ethnographic methods are helpful in generating rich and detailed accounts of software project teams, their interactions with project stakeholders, and their approaches for delivering products, as well as in-depth accounts of their experiences [24]. Hence, we would like to adopt the approach to leverage all contacts and insights we have from the cases.

### 3.1 Case Description

A case was selected from our convenient sample. We defined four criteria for our case selection: (1) a startup that operates for at least six months, such that their experience can be relevant, (2) a startup that has at least a first running prototype, (3) a startup that has at least an initial customer set, first customer payments or a group of users, (4) a startup that has software as core value of their business. We eventually decided to study the hypothesis-driven journey of two startup cases: case 1: Startuppuccino and case 2: MUML AS.

**Case 1** The startup is named after the name of the developed application, Star- tuppuccino [25], which is based at the Free University of Bozen-Bolzano in the northern part of Italy. Startuppuccino started with the experience and observation of two team members who are also university teachers. The initial idea of the teachers was to recommend good software tools to initiate and support startups that miss key skills in their teams (e.g., design, web development) [26]. Commonly, early-stage startups lack resources and look for some startup tools in order to launch their idea and test the product solution fit. Later, the idea pivoted into an educational platform that aims at helping entrepreneurship educators in providing students with better learning experience during their courses. Tools were also recommended to users at this level. So far the journey of Star- tuppuccino did three pivots: 1.) startuptools.club, 2.) MineToolz and 3.) current version running as Startuppuccino [25].

**Case 2** MUML AS is a spin-off from a Norwegian social media company. The CEO of the company quit the job and sought for a technical team to develop a hyper-local news platform. She started with the business idea and hiring several consultants, freelancers and contractors to realize and refine the idea. After that, a CTO joined the team and started a prototyping contract with a Vietnamese outsourcing team. The team was selected after a bidding process to en- sure the lowest price quote. The contract was made based on six-milestone delivery and payments were made after each milestone. The outsourcing team worked in a Sprint-based approach adopting Sprint planning and retrospective meetings, burn-down chart and communication via social media. After nine months of collaboration, the CEO stated that it was a positive experience regarding the value perceived. The outsourced team was offered to be a part of the startup.

## 3.2 Data Collection

Semi-structured individual interviews [27] and participant observation were used to collect data since they enable enough focus on the topic of interest, but also flexible structures to discover unforeseen information. Table 1 shows outlook of the data collection instrument. An interview guide was slightly different between two cases, between different people in the same case and even between the same interviewee subject. However, we asked three types of questions: (1) warm-up question about the current context of the interviewees related to business and product development, (2) past experience question to investigate how the interviewees did in certain project scenarios in the past and (3) lessons learnt questions to capture the beliefs that emerged or evolved from the project experiences. Most of our performed observations are active participation, in which researchers are members of the startups, actively involving in business development, decision making, product development and customer interaction. When counting observations with predefined research goals, there were six planned observation sessions conducted in MUML AS and ten planned observation sessions were conducted in Startuppuccino. The researchers came to observed sessions with a clear research goal in mind, sometimes with a check-list. Field note was done after the observation. In case of Startuppuccino, the observation of actions and thoughts were captured in a startup diary. Data triangulation was done by looking at project's artefacts, such as project plan, meeting notes, technical document and project management board. By triangulating our data sources and our instruments, we addressed issues of validity and obtained comprehensive insights into the application of ethnographic methods.

Table 1. Data collection instrument

| Cases | Data collection | Amount |
|---|---|---|
| Startuppuccino | | |
| | Planned participant observation-strategic meetings | 10 |
| | Interview with entrepreneur | 4 |
| | Artefacts: Trello, pitching videos, dairy project plan, project charts, kanban board | Various |
| MUML AS | | |
| | Planned participant observation-strategic meetings | 6 |
| | Interviews with entrepreneur | 3 |
| | Artefacts: pitching documents, trello, bitbucket, user research, project plan, Development contract | Various |

## 3.3 Data Analysis

Interview transcripts and observation diary were available for analysis. We adopted a narrative analysis by going through the scripts, identifying the relevant piece of text and labelled them by codes representing: business, product ideas and descriptions of MVP. Combining with extra materials, we came up with a list of hypotheses and MVPs. Hypotheses were either directly stated or indirectly explained by an interviewee. We also noted the timestamps when a hypothesis or an MVP occurs. The connections among hypotheses are interpretative and conducted by all co-authors of the work. For instance, the connection between hypotheses is interpreted by their semantic meanings. Most of the connections between hypotheses and MVPs are evident from our data. After that, a cross- case analysis was done to identify commonality and difference between two cases. This was done on top of the previous analysis of hypotheses and MVPs in each case.

Table 2. Hypotheses formulated in Startuppuccino journey

| ID  | Parent     | Hypothesis                                                                                                   | Tested-In    |
|-----|------------|--------------------------------------------------------------------------------------------------------------|--------------|
| H01 |            | Entrepreneurs have less time and resources to build startup so they need assistance from startup tools       |              |
| H02 |            | Entrepreneurs need right startup tool at right time for very early stage startups                            | M02          |
| H03 | H02        | People would like to see video on platform, Users like to grasp the idea quickly                             | M02, M03     |
| H04 | H02, H03   | People prefer a video with real users stating the idea                                                       | M02, M04     |
| H05 |            | Entrepreneurs/students need a better platform with guidance from mentors to intiate/run the startup          | M05          |
| H06 | H05        | Users like to grasp the idea quickly                                                                         | M05          |
| H07 |            | Students could know better about the startup course, Educators could get support to run the startup course   | M07          |

## 4 Results

This section describes our finding with regards to each case. First, we explain the Startuppuccino and then the MUML AS journey with the list of hypotheses formulated, then the MVPs that were created, the pivots that occurred and finally the relationship diagram between hypotheses and MVPs.

**4.1. Entrepreneurial Journey of Startuppuccino**

With regards to RQ1, we found that in Startuppuccino entrepreneurs had some initial ideas and assumptions about customer problems. Table 2 shows that most of the hypotheses relate to the customer problems, which is based on their business model canvas. Some hypothesis, for example, H04, was derived after obtaining the new knowledge from testing a previous hypothesis, i.e H02 and H03. Hence, we formulated a parent-child relationship between these hypotheses. The hypotheses are also temporally ordered; H01 is the first hypothesis and H07 is the last hypothesis in the investigated time-frame. During the postmortem analysis, we were also able to identify the MVPs that are associated with these hypotheses, as described in Table 3. We identify 7 MVPs (in which the pivots occurred at M02, M05, M07 as marked *) and 7 hypotheses as described in Table 3 and Table 2. MVPs were described with their types and how they were built in the startups. The MVP is numbered chronologically: M01 is the first MVP and M07 is the last one within our investigated time-frame. Pivots are evidence of visible knowledge and experience transfer in Startuppuccino. M02 is a zoom-in pivot, where a major change occurred in the team, targeted market, UX design of the product. M05 is a customer segment pivot, coming with new team members and vision change. M07 is the least knowledge transfer as it was a complete pivot, where the whole business model got changed.

### 4.2 Entrepreneurial Journey of MUML AS

With regards to RQ1, Table 4 shows that most of the hypotheses relate to the business objectives driven by their business model canvas. The hypotheses are also chronologically ordered; H01 is the first hypothesis and H14 is the last made so that users understand the idea quickly, made to retain users on the landing page made with real users at the startup weekend made with vision to provide support, entrepreneurs/students with startup tools and guidance provided by mentors made so that users understand the idea quickly made to retain users longer on platform made with vision changed to provide platform to support entrepreneurship education made with vision to support educators teaching course hypothesis in the investigated time-frame. During the postmortem analysis, we were also able to identify the MVPs that are associated with these hypotheses, as described in Table 5. We identify 13 MVPs (in which the pivots occurred at M03 and M13) and 14 hypotheses as described in Table 5 and Table 4. MVPs were described with their types and how they were built in the startups. The MVP is numbered chronologically: M01 is the first MVP and M13 is the last one within our investigated time-frame. In MUML AS, two pivots happen, which occurred by building a new MVP (M3 and M13) based on previous learning from customer needs and product design. M03 is a customer need pivot, which is quite disconnected from its previous MVP. However, the learning experience regards to UX design and customer involvement remained the same with those in previous MVPs. M13 is a technology pivot, where new market research results in a new technical platform. Only the platform was changed here, all the knowledge about the customer, product design and business model remained the same.

Table 3. MVPs build in Startuppuccino journey MVP

| ID | MVP | Description |
|---|---|---|
| M01 | Mockup | made to visualize, understand the very first idea clearly |
| M02* | Landing page | made just enough for the market/users |
| M03 | Explainer Video | made so that users understand the idea quickly, made to retain users on the landing page |
| M04 | Explainer video | made with real users at the startup weekend |
| M05* | Concierge | made with vision to provide support, entrepreneurs/students with startup tools and guidance provided by mentors |
| M06 | Explainer Video | made so that users understand the idea quickly made to retain users longer on platform |
| M07* | Concierge | made with vision changed to provide platform to support entrepreneurship education made with vision to support educators teaching course |

**4.3 Findings from cross-case analysis**

We observe some commonalities in terms of hypothesis and MVP development in the two startup cases: With regards to RQ1, we found that startups do actually learn during entrepreneurial evolution and the learning can be marked with either hypothesis testing or MVP creation. However, the overall learning does not occur systematically and linearly. The relationship between hypotheses and MVPs is non-linear. The theoretical model of startup evolution includes a series of incremental experiments that involves hypothesis testing. In both cases, we find that the actual model of hypothesis testing in startups is more complicated. It is not straightforward that a hypothesis is associated with an MVP. In some cases, a business hypothesis is tested by multiple MVPs, at different times in the startup life-cycle. Validating one hypothesis can lead to another hypothesis (parent-child relationship). In some cases, one hypothesis can be derived from multiple parent hypothesis. Some hypotheses are so complex that they are fully tested by the very late MVPs. We also observe some MVPs that answer multiple hypotheses. These are often important MVPs that turn into commercial prod- ucts. With regards to RQ2, We capture the relationship between hypotheses and MVPs as in the Figure 4. In the figure, the dashed link represents the temporal relationship or the evolution flow over time of the startup. The white-head arrow links represent the parent-child relationship of the hypotheses. The black-head arrow links represent the evolution of MVPs. It is also used for the association link between a hypothesis and an MVP. In the case M1 there is no link with the hypothesis as the MVP was never validated. In the case of M2, which was built on top of M1, the pivot occurred hence it is highlighted green. In reference to the case of M5 and M6, the pivot occurred during M5, but M6 was tested at the same time of M5. Both MVPs were developed in parallel around the same time. In relation to RQ2, we found that there are NO correspondences between hy- potheses and MVPs. According to Lean Startup, learning occurs while validating pre-defined hypotheses. However, we find in both cases that some MVP is built without an association to a hypothesis. The MVP is built either as an extension of a previous one or with the push from customer and market demands. There are also hypotheses not tested. Startup founders recognize that derived hypothe- ses were not fully covered by MVPs. Some are skipped due to intuitive reasons; some are skipped mistakenly. Moreover, we found that pivot can be captured from the MVP-creation approach. A pivot marked by a new MVP often inherits learning from the previous MVPs. Typically, the pivoted MVP will start from scratch. This means an MVP before the pivoted one, is typically considered as a throw-away prototype. There are also situations in which a pivoted MVP reuses source code from the previous MVPs. In our cases, the reuse also involves a significant refactoring and change of code bases. A pivoted MVP is also found to be associated with a new (sub) hypothesis disconnected with the previous hypothesis.

Table 4. Hypotheses formulated in MUML AS

| ID | Parent | Hypothesis | Tested-In |
|---|---|---|---|
| H01 | | People are interested in hyper-local news around them | M01 M02,M12 |
| H02 | H01 | People are interested in a sub-set of news depending on geographical context | |
| H03 | H01 | People are interested in trusted, validated news | M09 |
| H04 | H01, H02 | People are interested in news in other locations as well | M07 |
| H05 | | People are willing to share hyper-local news around them | M01, M12 |
| H06 | H05 | People are interested in sharing news via interesting sharing mechanism | M09, M10 |
| H07 | | People are interested in news displayed in a map | M04 |
| H08 | H07 | People like to see news headline in the map | M04, M06 |
| H09 | H07 | People like to be able to configure the radius of news they can receive | M08 |
| H10 | H01, H07 | People would like to see picture and less text | M04, M06,M07 |
| H11 | H10 | People would like to see picture, live stream video as well | M10 |
| H12 | H05 | There is a way to trigger people to post news | M09 |
| H13 | H12 | A camera-ready button triggers the willingness to capture a photo and share | M09 |
| H14 | H12 | Gamification can help users to engaged into the system | M12 |

Table 5. MVPs build in MUML AS journey

| ID | MVP | Description |
|---|---|---|
| M01 | Explainer Video | Firstly made to express the business idea |
| M02 | Mockup | Created by a consultant company to communicate ideas |
| M03* | Mockup | Created in just in mind, use to communicate the idea with CEO, with designers and development team |
| M04 | Single feature | The first implemented features include Mapview, Listview of a news |
| M05 | Single feature | No new feature added but changing a lot relating to the interfaces |
| M06 | Evolutionary | Adding detail view, location, features to the app |
| M07 | Evolutionary | Channel feature |
| M08 | Evolutionary | Map configuration feature |
| M09 | Evolutionary | Camera button |
| M10 | Evolutionary | Live story feature, preparing for two pitching events and a makerfaire |
| M11 | Landing page | Formal page of the startup |
| M12 | Evolutionary | User management and gamification feature |
| M13* | Evolutionary | Making the new version of MUML AS for Android devices It was previously applied for iphone only. |

## 5 Discussion and Conclusions

This study describes the hypothesis-driven journey of two software startups expedition which started with forming the hypothesis, building MVPs and pivots that occurred. Lean Startup and previous studies on software startups have neglected the relationship between hypothesis and MVPs or considered them in an ideal context. We found that entrepreneur does learn from testing their hypotheses, however, they do not always focus on hypothesis formulation and hence, the relationship between business objective to test and MVPs to build is not always straightforward.

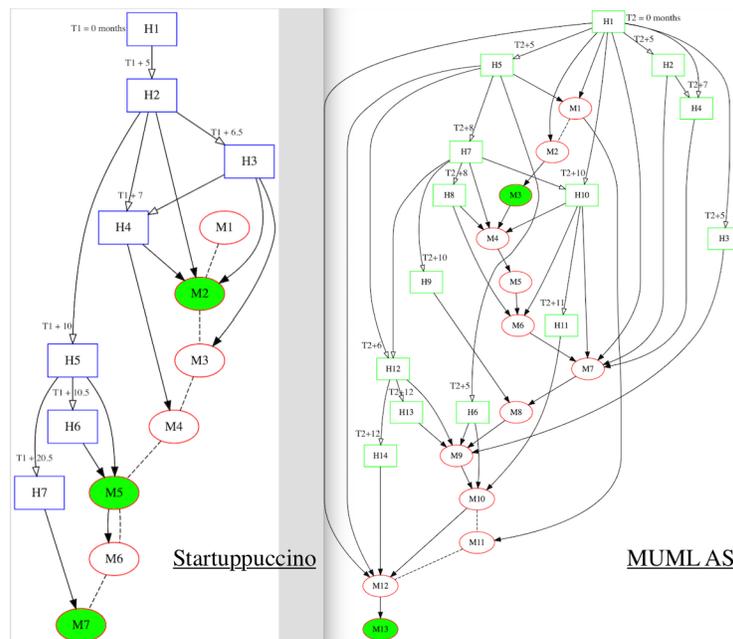

Fig.4. Relationship between Hypotheses and MVPs in Startuppuccino and MUML AS

Through two case studies, we observed that a relationship be- tween hypothesis and MVPs is non-linear and incomplete. We also proposed an approach to visualize the startup journey from capturing the Hypothesis-MVP relationships. From our cases, it seems that the amount of learning entrepreneur have depends on user involvement and their existing knowledge about market, industry and technology. Little user involvement might lead to little experience gained from testing hypotheses. For an entrepreneur, it is crucial to solving the urgent problem of a user, although a startup has to face a complete pivot. This could be time-consuming and a big move for a startup to deal with, but benefi- cial too. Moreover, an entrepreneur should grab every opportunity to experiment with MVPs. Furthermore, the need and effectiveness of having a strong business driver for a startup are important. Last but not least, with the usefulness of visualizing startup journeys demonstrated in this paper, an entrepreneur can find the journey maps an useful tool for reflecting and reviewing possible gaps in the business and product development. We are not aware of a specific toolset for this purpose in the market. However, an entrepreneur can use generic graph tools, such as Graphviz, GraphTea and Plotly and follow the approach described in this paper. There are several threats to validity worth to discuss [28]. One internal threat of validity is the bias in data collection, as data might not represent a comprehensive story. In order to mitigate this threat, we selected CEOs during the postmortem analysis, who have the best understanding about their startups. We used all opportunities for interviewing relevant people of our cases in this context of the study. We also used artefacts (Trello, project charts, kanban board, dairy) during postmortem to increase our understanding of the cases. With both startups, we also acted as startup team members, which enables a lot of insights beyond interviews. Another internal threat to validity regards how reliable the reported cases are. This ensured that all of the authors have not only theoretical background about software startups but also hands-on experience. A construct threat to validity is a possible inadequate description of constructs. An external threat to validity is the representativeness of our selected cases. Both of the cases are small startups. Besides, the startup decisions on MVP might be influenced by individual personalities. Future research can validate results from this work by systematic adoption of the approach in a larger set of cases. We also call for a development of a specific toolset to visualize startups hypotheses, MVPs, and the connections among them. The toolset will definitely highlight the learning and experience flow during the entrepreneurial development.

## References

1. Fletcher, D.E.: Entrepreneurial processes and the social construction of opportu- nity. Entrepreneurship and Regional Development. 18(5), 21–440 (2006)
2. Song, M., Podoynitsyna, K., Van Der Bij, H., Halman, J.I.: Success factors in new ventures: A metaanalysis. Journal of product innovation management. 25(1), 7–27 (2008)
3. Duc, A.N., Abrahamsson, P.: Minimum viable product or multiple facet product? The Role of MVP in software startups. In International Conference on Agile Soft- ware Development. 118–130 (2016)
4. Bajwa, S.S., Wang, X., Duc, A.N., Abrahamsson, P.: Failures to be celebrated: an analysis of major pivots of software startups. Empirical Software Engineering. 22(5), 2373–2408 (2017)
5. Sarasvathy, S.D.: Effectuation: Elements of entrepreneurial expertise. Edward El- gar Publishing. (2009)
6. Venkataraman, S., Sarasvathy, S.D., Dew, N., Forster, W.R.: Reflections on the 2010 AMR decade award: Whither the promise? Moving forward with en- trepreneurship as a science of the artificial. Academy of Management Review. 37(1), 21–33 (2012)
7. Lichtenstein, B.B.: Generative emergence: A new discipline of organizational, en- trepreneurial, and social innovation. Oxford University Press (UK). (2014)
8. Ries, E.: The lean startup: How today's entrepreneurs use continuous innovation to create radically successful businesses. Crown Books. (2011)
9. Mu˙ller, R.M., Thoring, K.: Design thinking vs. lean startup: A comparison of two user-driven innovation strategies. Leading through design. 151 (2012)
10. Khanna, D., 2018: Experiential Team Learning in Software Startups. In Interna- tional

   Conference on Agile Software Development. Springer, Cham (2018)
11. Basili, V.R., Selby, R.W., Hutchens, D.H.: Experimentation in software engineer- ing. IEEE Transactions on software engineering. (7), 733–743 (1986)
12. Blank, S.: The four steps to the epiphany: successful strategies for products that win. BookBaby. (2013)
13. Eisenmann, T., Ries, E., Dillard, S.: Hypothesis-driven entrepreneurship: the lean startup. Harvard Business School Entrepreneurial Management Case No. 812–095 (2012)
14. Wang, X., Khanna, D., Abrahamsson, P.: Teaching Lean Startup at University: An Experience Report. International Workshop on Software Startups (IWSS) co- located with 22nd ICE/IEEE International Technology Management Conference. (2016)
15. Selden, P.D., Fletcher, D.E.: The entrepreneurial journey as an emergent hierar- chical system of artifact-creating processes. Journal of Business Venturing. 30(4), 603–615 (2015)
16. Simon, H.A.: The sciences of the artificial. MIT press. (1996)
17. Nguyen Duc, A. Wang X., Abrahamsson, P.: What Influences the Speed of Pro- totyping? An Empirical Investigation of Twenty Software Startups, Norwegian.  (2017)
18. Lindgren, E., Münch, J.: Raising the odds of success: the current state of ex- perimentation in product development. Information and Software Technology. 77, 80–91 (2016)
19. Fagerholm, F., Guinea, A.S., Mäenpää, H. and Münch, J.: Building blocks for continuous experimentation. In Proceedings of the 1st international workshop on rapid continuous software engineering. ACM. 26–35 (2014)
20. Nguyen Duc, A., Seppänen, P., Abrahamsson, P.: Hunter-gatherer cycle: a concep- tual model of the evolution of startup innovation and engineering. 1st Workshop on Open Innovation on Software Engineering, ICSSP. (2015)
21. Sharp, H., Dittrich, Y., De Souza, C.R.: The role of ethnographic studies in em- pirical software engineering. IEEE Transactions on Software Engineering. 42(8), 786–804 (2016)
22. Fetterman, D.M.: Ethnography: Step-by-step Sage. 17, (2010)
23. Reeves, S., Kuper, A., Hodges, B.D.: Qualitative research methodologies: ethnog- raphy. BMJ: British Medical Journal. 337, (2008)
24. Passos, C., Cruzes, D. S., Dybå, T., Mendonça, M.: Challenges of applying ethnog- raphy to study software practices. In Empirical Software Engineering and Measure- ment (ESEM), 2012 ACM-IEEE International Symposium on IEEE. 9–18 (2012)
25. Khanna, D., Mondini, M., Pantiuchina, J., Stillittano, G., Wang, X.: Experi- ment with MVPs: the First Startuppuccino Steps to a Lean Edtech Startup. In agilealliance, retrive from https://www.agilealliance.org/resources/experience- reports/experiment-with-mvps/. (2017)
26. Edison, H., Khanna, D., Bajwa, S. S., Brancaleoni, V., Bellettati, L.: Towards a Software Tool Portal to Support Startup Process. In International Conference on Product-Focused Software Process Improvement, Springer International Publish- ing. 577–583 (2015)
27. Myers, M.D. and Newman, M.: The qualitative interview in IS research: Examining the

craft. Information and organization. 17(1), 2–26 (2007)
28. Runeson, P., Höst, M.: Guidelines for conducting and reporting case study research in software engineering. Empirical software engineering. 14(2), 131 (2009)